\documentstyle[preprint,prd,epsfig,aps,12pt]{revtex}
\def\be{\begin{equation}}
\def\ee{\end{equation}}
\def\bea{\begin{eqnarray}}
\def\eea{\end{eqnarray}}

\begin{document} 
\thispagestyle{empty}
\preprint{\vbox{
\halign{&##\hfil\cr
        & DESY 99-061 \cr
        & PUTP-98-35 \cr
        & {} \cr
        & {} \cr }}}
\title {\Large Possible retardation effects of quark confinement on the
meson spectrum II\\[5mm]}
\author{Cong-Feng Qiao\footnote{Alexander von Humboldt fellow}\\[2mm]}
\address {CCAST (World Laboratory), P.O. Box 8730, Beijing 100080, P.R.
China \\ and II Institut f\"ur Theoretische Physik, Universit\"at
Hamburg, 22761 Hamburg, Germany}
\vskip -8mm
\author{Han-Wen Huang$^\ast$ \\[2mm]}
\address{Fakult\"at f\"ur Physik, Universit\"at Bielefeld, 33615 
Bielefeld, Germany} 
\vskip -8mm
\author{Kuang-Ta Chao \\[2mm]}
\address {CCAST (World Laboratory), P.O. Box 8730, Beijing 100080, P.R.
China \\ and Department of Physics, Peking University, 
Beijing 100871, P.R. China}
\maketitle
\vspace{0.6cm}
\begin{abstract}
\vskip -12mm
\noindent
We present the results of a study of heavy-light-quark bound states
in the context of the reduced Bethe-Salpeter equation with relativistic
vector and scalar interactions. We find that satisfactory fits may also 
be obtained when the retarded effect of the quark-antiquark interaction 
is concerned.
\vskip 0.8cm
\noindent
PACS number(s): 12.39.Ki, 12.39.Pn, 12.40.Yx
\end{abstract} 
\newpage
Because of the limited understanding for confinement at present, more 
theoretical efforts to be made related to this issue are worthwhile. In an
earlier paper \cite{a1}, we presented results of a relativistic analysis
of the spectrum of light- and heavy-quark-antiquark bound states based on
the reduced Bethe-Salpeter(BS) equation, while retardation effects in the
quark-antiquark interaction kernel were taking into consideration. 
The results are stimulating and appears having clarified the problem
pointed out by Durand {\it et al.} \cite{durand} for the static scalar
confinement in reduced Salpeter equation. The "intrinsic flaw" of the
Salpeter equation with static scalar confinement could  be remedied to
some extent by taking the retardation effect into confinement. In the
on-shell approximation for the retardation term of linear confinement,
the notorious trend of narrow level spacings for quarkonium states
especially for light quarkonium states is found to be removed. A good fit
for mass spectrum of S-wave heavy and light quarkonium states (except for
the light pseudoscalar mesons) is obtained using one-gluon exchange
potential and the scalar linear confinement potential with retardation
taken into account. 

In this paper we extend our previous study to the heavy-light-quark
systems ($Q\bar{q}$ or $q\bar{Q}$) in order to get a fully understanding
of the retardation effects on meson spectra. Based on the same procedure
taken in Ref.\cite{a1}, by solving the reduced Bathe-Salpeter equation
numerically, the previous conclusion is further substantiated through
this study.

We assume the confinement kernel in momentum space taking the form
\begin{equation}
\label{2}
G(p)\propto \frac{1}{(-p^2)^2}=\frac{1}{(\vec{p}^2 - p_0^2)^2},
\end{equation}
as suggested by some authors as the dressed gluon propagator to implement
quark confinement \cite{pa}. Here $p$ is the $4$-momentum exchanged
between the quark and antiquark in a meson. If the system is not highly
relativistic we may make the approximation
\begin{equation}
\label{3} 
G(p)\propto\frac{1}{(\vec{p}^2 - p_0^2)^2}\approx\frac{1}{(\vec{p}^2)^2}
\left(1 + \frac{2 p_0^2}{\vec{p}^2}\right),
\end{equation}
and may further express $p_0$ in terms of its on-shell values which are 
obtained by assuming that quarks are on their mass shells. This should be a
good approximation for $c\bar{c}$ and  $b\bar{b}$ systems. However, to get a
qualitative feeling about the retardation effect considered here, we will
also use (\ref{3}) for heavy-light-quark mesons, though the
approximations are not as good as for heavy-heavy-quark mesons. With
the above  procedure, the scalar confinement kernel becomes instantaneous
again but with some retardation effects have been taken into the kernel.
In the static limit, the retardation term vanishes and the kernel returns
to $G(q)\propto \frac{1}{(\vec{p}^2)^2}$, which is just the Fourier
transformation of the linear confining potential. In this paper, we will
use this modified scalar confining potential in which the retardation
effect is incorporated and the one-gluon-exchange potential in the
framework of the reduced Salpeter equation to study the mass spectrum of
$q\bar{Q}$ mesons, and the structure of the hyperfine splittings of the
heavy-light-quark systems will also be investigated in this paper.

In quantum field theory, one of the basic descriptions for the bound
states is the Bethe-Salpeter equation \cite{s5}. We can define the
BS wave function of the bound state $\mid P\rangle $ of a quark
$\psi(x_1)$ and an antiquark $\bar{\psi}(x_2)$ as
\begin{equation}             
\label{e1}
\chi(x_1,x_2)=\langle 0\mid T\psi(x_1)\bar{\psi}(x_2)\mid P\rangle.
\end{equation}
Here $T$ represents the time-order product, and the wave function can be
transformed into the momentum space,
\begin{equation}
\label{e2}
\chi_P(q)=e^{{-iP}\cdot X}\int d^{4}x e^{{-iq}\cdot x} \chi(x_1,x_2),         
\end{equation}
where $P$ is the four-momentum of the meson and $q$ is the relative
momentum between quark and antiquark. As the standard measure in solving
the BS equation, we choose center-mass and relative coordinates as 
variables,
\begin{equation}
X = \eta_1 x_1 + \eta_2 x_2,~~~x=x_1 - x_2,
\end{equation}
where $\eta_i=\frac{m_i}{m_1+m_2}~(i=1,2)$. After taking the Fourier
transformation, in the momentum space the BS equation reads as
\begin{equation}
\label{e3}
(\not\!{p_1}-m_1)\chi_P(q)(\not\!{p}_2+m_2)=\frac{i}{2\pi}
\int d^{4}k G(P,q-k)\chi_P(k), 
\end{equation}
where $ p_1 $ and $ p_2 $ represent the momenta of quark and antiquark,
respectively,
\begin{equation}
\label{e4}
p_1=\eta_1 P + q,~~~  p_2=\eta_2 P - q, 
\end{equation}
and $G(P,q-k)$ is the interaction kernel which acts on $ \chi $ and is
determined by the interquark dynamics. Note that in Eq.(\ref{e3}) $m_1$
and $m_2$ represent the effective constituent quark masses so that we
could use the effective free propagators of quarks instead of the full 
propagators. This is an important approximation and simplification for
light quarks. Furthermore, because of the lack of a fundamental
description for the nonperturbative QCD dynamics, we have to make some
approximations for the interaction kernel of quarks. In solving the 
Eq.(\ref{e3}), the kernel is taken to be instantaneous, but with some
retardation effect being taken into it; the negative energy projectors in
the quark propagators are neglected, because in general the negative
energy projectors only contribute in higher orders due to $M - E_1 - E_2
\ll M + E_1 + E_2 $, where M, $E_1$, and $E_2$ are the meson mass, the
quark kinetic energy, and the antiquark kinetic energy respectively.
Based on the above assumptions the BS equation can be reduced to a
three-dimensional equation, i.e. the reduced Salpeter equation,
\begin{equation}
\label{e11}
(P^0-E_1-E_2)\Phi_{\vec P}(\vec q)=
\Lambda_{+}^{1}\gamma^0\int d^3 k G(\vec P,\vec q ,\vec k)\Phi_{\vec P}(\vec k)
\gamma^0 \Lambda_{-}^{2}.
\end{equation}
Here
\begin{equation}
\label{e10}
\Phi_{\vec P}(\vec q)=\int dq^0 \chi_P(q^0,\vec q),
\end{equation}
is the three dimensional BS wave function, and
\begin{eqnarray}
\label{e12}
\Lambda_{+}^{1}={1\over 2E_1}(E_1+\gamma^0 {\vec {\gamma}}
    \cdot{\vec p_1}+m_1 \gamma^0 ),\\ 
\Lambda_{-}^{2}={1\over 2E_2}(E_2-\gamma^0 {\vec {\gamma}}
\cdot{\vec p_2}-m_2 \gamma^0 ),
\end{eqnarray}
are the remaining positive energy projectors of the quark and antiquark
respectively, with $E_1 = \sqrt{m_1^2+\vec{p_1}^2},~E_2 =\sqrt{m_2^2 + 
\vec{p_2}^2}$. The formal products of $G\Phi$ in Eq.(\ref{e11}) take
the form

\begin{equation}
\label{e13}
G\Phi=\sum\limits_{i} G_i O_i \Phi O_i = G_s \Phi + \gamma_{\mu}\otimes
   \gamma^{\mu} G_v \Phi, 
\end{equation}
where $ O_i=\gamma_{\mu} $ corresponding to the perturbative  
one-gluon-exchange interaction and $O_i=1$ to the scalar confinement 
potential.
  
From Eq.(\ref{e11}) it is easy to see that
\begin{equation}
\label{e14}
\Lambda_{+}^{1}\Phi_{\vec P}(\vec q)=\Phi_{\vec P}(\vec q),~ 
\Phi_{\vec P}(\vec q)\Lambda_{-}^{2}=\Phi_{\vec P}(\vec q).
\end{equation}
Considering the constraint of Eq.(\ref{e14}) and the requirements of
space reflection of bound states, in the meson rest frame 
$({\stackrel{\rightharpoonup }{P}}=0)$ the wave function 
$\Phi_{\vec P}(\vec q) $ for the $0^-$ and $1^-$ mesons can be expressed 
as 
\begin{eqnarray}
\label{a31}
\Phi _{\stackrel{\rightharpoonup }{P}}^{0^{-}}(\stackrel{\rightharpoonup}
{q}) =\Lambda _{+}^1\gamma^0(1+\gamma^0)\gamma_5\gamma^0\Lambda_{-}^2
\varphi(\stackrel{\rightharpoonup}{q}),
\end{eqnarray}
\begin{eqnarray}
\label{a31aa}
\Phi_{\stackrel{\rightharpoonup }{P}}^{1^{-}}( \stackrel{\rightharpoonup}
{q})=\Lambda _{+}^1\gamma^0(1+\gamma^0)\not\!e\gamma^0\Lambda_{-}^2f
(\stackrel{\rightharpoonup}{q}),
\end{eqnarray}
where $\not\!e =\gamma_{\mu} e^\mu$ is the polarization vector of   
$1^-$ meson; $\varphi(\vec{q})$ and $f(\vec{q})$ are scalar functions
of $\vec{q}^2$. It is easy to show that Eqs.(\ref{a31}) and (\ref{a31aa})
are the most general forms of the S-wave wave functions for $0^-$ and
$1^-$ mesons in the rest frame.

Substituting Eqs.(\ref{e13}), (\ref{a31}), and (\ref{a31aa}) into
Eq.(\ref{e11}), one can derives the equations out for 
$\varphi(\stackrel{\rightharpoonup}{q})$ and $f(\stackrel{\rightharpoonup}
{q})$ in the meson rest frame \cite{t9},
\begin{eqnarray}
\label{a26}
M\varphi _1(\stackrel{\rightharpoonup }{q})
&=&(E_1+E_2)\varphi _1(\stackrel{\rightharpoonup }{q})\nonumber 
\\
&&-\frac{E_1E_2+m_1m_2+\stackrel{\rightharpoonup }{q}^2}{4E_1E_2}
\int d^3k(G_S(\stackrel{\rightharpoonup}{q}, \stackrel{\rightharpoonup}
{k})-4G_V(\stackrel{\rightharpoonup}{q}, \stackrel{\rightharpoonup }
{k}))\varphi _1(\stackrel{\rightharpoonup}{k})\nonumber \\  
&&-\frac{(E_1m_2+E_2m_1)}{4E_1E_2}\int d^3k(G_S(\stackrel{\rightharpoonup}
{q}, \stackrel{\rightharpoonup }{k})+2G_V(\stackrel{\rightharpoonup}{q}, 
\stackrel{\rightharpoonup }{k}))\frac{m_1+m_2}{E_1+E_2}\varphi_1(\stackrel
{\rightharpoonup}{k})\nonumber\\  
&&+\frac{E_1+E_2}{4E_1E_2}\int d^3kG_S(\stackrel{\rightharpoonup}{q}, 
\stackrel{\rightharpoonup}{k})(\stackrel{\rightharpoonup }{q}\cdot 
\stackrel{\rightharpoonup }{k})\frac{m_1+m_2}{E_1m_2+E_2m_1}
\varphi_1(\stackrel{\rightharpoonup }{k})\nonumber \\  
&&+\frac{m_1-m_2}{4E_1E_2}\int d^3k(G_S(\stackrel{\rightharpoonup }{q}, 
\stackrel{\rightharpoonup}{k})+2G_V(\stackrel{\rightharpoonup}{q}, 
\stackrel{\rightharpoonup }{k}))(\stackrel{\rightharpoonup}{q}\cdot
\stackrel{\rightharpoonup}{k})\frac{E_1-E_2}{E_1m_2+E_2m_1}
\varphi_1(\stackrel{\rightharpoonup }{k}),~~~
\end{eqnarray}
with
\begin{equation}
\label{a27}
\varphi _1(\stackrel{\rightharpoonup}{q}) = \frac{(m_1+m_2+E_1+E_2)
(E_1m_2+E_2m_1)}{4E_1E_2(m_1+m_2)}\varphi (\stackrel{\rightharpoonup}{q}),
\end{equation}
and
\begin{eqnarray}
\label{a28}
Mf_1(\stackrel{\rightharpoonup}{q})
&=&(E_1+E_2)f_1(\stackrel{\rightharpoonup }{q})\nonumber 
\\
&&-\frac 1{4E_1E_2}\int d^3k(G_S(\stackrel{\rightharpoonup }{q}, 
\stackrel{\rightharpoonup }{k})- 2G_V(\stackrel{\rightharpoonup}{q}, 
\stackrel{\rightharpoonup }{k}))(E_1m_2+E_2m_1)f_1(\stackrel
{\rightharpoonup }{k})\nonumber \\  
&&
-\frac{E_1+E_2}{4E_1E_2}\int 
d^3kG_S(\stackrel{\rightharpoonup }{q}, \stackrel{\rightharpoonup}{k})
\frac{E_1m_2+E_2m_1}{E_1+E_2}f_1(\stackrel{\rightharpoonup }{k})\nonumber\\  
&&+\frac{E_1E_2-m_1m_2+\stackrel{\rightharpoonup}{q}^2}{4E_1E_2\stackrel
{\rightharpoonup }{q}^2} \int d^3k(G_S(\stackrel{\rightharpoonup }{q}, 
\stackrel{\rightharpoonup}{k})+4G_V(\stackrel{\rightharpoonup }{q}, 
\stackrel{\rightharpoonup}{k}))(\stackrel{\rightharpoonup }{q}
\cdot\stackrel{\rightharpoonup }{k})f_1(\stackrel{\rightharpoonup }
{k})\nonumber \\  
&&-\frac{E_1m_2-E_2m_1}{4E_1E_2\stackrel{\rightharpoonup}{q}^2} 
\int d^3k(G_S(\stackrel{\rightharpoonup}{q}, \stackrel{\rightharpoonup}
{k})-2G_V(\stackrel{\rightharpoonup}{q}, \stackrel{\rightharpoonup }{k}))
(\stackrel{\rightharpoonup }{q}\cdot\stackrel{\rightharpoonup 
}{k})\frac{E_1-E_2}{m_2+m_1}f_1(\stackrel{\rightharpoonup }{k})\nonumber\\ 
&&-\frac{E_1+E_2-m_2-m_1}{2E_1E_2\stackrel{\rightharpoonup}{q}^2} \int 
d^3kG_S(\stackrel{\rightharpoonup }{q}, \stackrel{\rightharpoonup }{k})
(\stackrel{\rightharpoonup }{q}\cdot\stackrel{\rightharpoonup }{k})^2
\frac 1{E_1+E_2+m_1+m_2}f_1(\stackrel{\rightharpoonup}{k})\nonumber \\ 
&&-\frac{m_2+m_1}{E_1E_2\stackrel{\rightharpoonup}{q}^2} \int
d^3kG_V(\stackrel{\rightharpoonup }{q}, \stackrel{\rightharpoonup}{k})
(\stackrel{\rightharpoonup }{q}\cdot\stackrel{\rightharpoonup }{k})^2
\frac 1{E_1+E_2+m_1+m_2}f_1(\stackrel{\rightharpoonup }{k}),
\end{eqnarray}
with
\begin{equation}
\label{a29}
f_1(\stackrel{\rightharpoonup }{q})= -\frac{m_1+m_2+E_1+E_2}{4E_1E_2}
f(\stackrel{\rightharpoonup }{q}).  
\end{equation}
Eqs.(\ref{a26}) and (\ref{a28}) can also be formally expressed as more
compact forms
\begin{eqnarray}
\label{e22}
(M-E_1-E_2)\varphi_{1}({\vec q})=\int d^{3} k\sum\limits_{{i=S},{V}}
F_{i}^{0^-} ({\vec q},{\vec k})G_i({\vec q},{\vec k})\varphi_{1}({\vec k}),
\end{eqnarray}
\begin{eqnarray}
\label{e22aa}
(M-E_1-E_2)f_{1}({\vec q})=\int d^{3} k\sum\limits_{{i=S},{V}}
F_{i}^{1^-} ({\vec q},{\vec k})G_i({\vec q},{\vec k})f_{1}({\vec k}).
\end{eqnarray}
In case of taking the nonrelativistic limit for both quark and antiquark,
expanding in terms of ${\stackrel{\rightharpoonup }{q}}^2/{m_1}^2$ and
${\stackrel{\rightharpoonup }{q}}^2/{m_2}^2$, it can be approved that
Eqs.(\ref{a26}) and (\ref{a28}) are identical with the Schr\"odinger
equation to the zeroth order, and with the Breit equation to the first
order. 

To solve Eq.(\ref{e3}) one must have a good command of the potential
between two quarks. At present, the reliable information about the 
potential only comes from the lattice QCD result, which shows that the
potential for a heavy quark-antiquark pair $Q\bar{Q}$ in the static limit
is well described by a long-ranged linear confining potential ( Lorentz
scalar $V_S$ ) and a short-ranged one gluon exchange potential ( Lorentz
vector $V_V$ ), i.e. \cite{la,lb}
\begin{eqnarray}
\label{a9}
&&{V(r)}={V_S(r)+\gamma_{\mu}\otimes\gamma^{\mu} V_V(r)},
\end{eqnarray}
with
\begin{eqnarray}
{V_S(r)}={\lambda r\frac {(1-e^{-\alpha r})}{\alpha r}},
\end{eqnarray}
\begin{eqnarray}
{V_V(r)}=-{\frac 43}{\frac {\alpha_{s}(r)} r}e^{-\alpha r},
\end{eqnarray}
where the introduction of the factor $e^{-\alpha r}$ is not only for the 
sake of avoiding the infrared(IR) divergence but also incorporating the
color screening effects of the dynamical light quark pairs on the
"quenched" $Q\bar{Q}$ potential \cite{t13}.  Although the lattice QCD
result for the $Q\bar{Q}$ potential is supported by the heavy quarkonium
spectroscopy including both spin-independent and spin-dependent
effects\cite{s1,s2,ding}. we will employ this static potential below to
the heavy-light-quark systems as an assumption. The interaction potentials
in Eq.(\ref{a9}) can be transformed straightforwardly into the momentum
space, where the strong coupling constant 
\begin{equation}
\label{a15}
\alpha _s( \stackrel{\rightharpoonup }{p}) =\frac{12\pi }{27}\frac 1{\ln
(a+\frac{\stackrel{\rightharpoonup }{p}^2}{\Lambda_{QCD}^2})}.
\end{equation}
is assumed to be a constant of $O(1)$ as 
${\stackrel{\rightharpoonup}{p}}^2\rightarrow 0$. The constants $\lambda$,
$\alpha$, $a$, and $\Lambda_{QCD}$ are the parameters that characterize
the potential.

In taking the retardation effect of scalar confinement into consideration, 
as discussed in Ref.\cite{a1}, the confinement will be approximately
introduced by adding a retardation term  $\frac{2 p_0^2}{\vec{p}^6}$ to
the instantaneous part $\frac{1}{(\vec{p}^2)^2} $ as given in
Eq.(\ref{3}),  and $p_0^2$ will be treated to take its on-shell values
which are obtained by assuming that the quarks are on their mass shells,
which means that the retardation term will become instantaneous rather
than convoluted. By this procedure the modified scalar confinement
potential will include retardation effect and become
\vskip 0.2cm
\begin{eqnarray}
\label{p24}
G_S(\vec{p})\rightarrow G_{S}(\vec{p},\vec{k})&=& 
-\frac{\lambda}{\alpha}\delta^3(\vec{p}) +
\frac{\lambda}{\pi^2}\frac{1}{(\vec{p}^2 + \alpha^2)^2} \nonumber \\
&+& \frac{2 \lambda}{\pi^2}\frac{1}{(\vec{p}^2 + \alpha^2)^3}
\big(\sqrt{(\vec{p}+\vec{k})^2+m^2} - \sqrt{\vec{k}^2+m^2}~\big)^2,
\end{eqnarray}
which is related not only to the interquark momentum exchange $\vec{p}$
but also the quark momentum $\vec{k}$ itself. 

Based on the formalism and discussions above, we can now embark on the
numerical calculations, in which we take the following values for input
parameters
\begin{eqnarray}
\label{a16}
\lambda=0.21~GeV^{2}, ~\alpha=0.06~GeV,
~a=e=2.7183,~\Lambda_{QCD}=0.19~GeV, ~C = -0.05 
\end{eqnarray}
and 
\begin{eqnarray}
\label{e24}
m_u=m_d=0.35~{\rm GeV},~ m_s=0.5~{\rm GeV},
m_c=1.68~{\rm GeV},~ m_b=4.925~{\rm GeV},
\end{eqnarray}
which fall in the scopes of customarily usage. The numerical results with
retardation are listed in Table I. For the convenience of comparisons,
results obtained without retardation  are also presented. 

From Table I one can immediately see that the calculated masses with
retardation effect are generally well fitted compared with those without
retardation considered, and the spin splittings, $1^3S_1 - 1^1S_0$ are
significantly improved by adding the retardation term to the scalar
confinement potential. These conclusions obviously shed light on the
usefulness of the Bethe-Salpeter equation in describing systems containing
light quarks, however, in the mean time they also ask for further
investigations on the interaction kernels.

As noticed in Ref.\cite{durand}, the smallness of the hyperfine splitting
obtained for $q\bar{Q}$ mesons is due to the weakness of the binding
potential in these systems. However, similarly as showed in Ref.\cite{a1}
this
situation may also be changed after including the retardation effect in
the interaction kernel. For demonstration, in the equal-quark-mass special
case the coefficients for the scalar potential $G_s$ which plays the main
role in setting the spin splittings in Eqs.(\ref{e22}) and (\ref{e22aa})
will reduce to 
\begin{eqnarray}
\label{bb1}
F_{S}^{0^-}(\vec{q},\vec{k})= -\frac{1}{2}+ 
\frac{\vec{q}\cdot\vec{k}-m^2}{2 E_q E_k},
\end{eqnarray}
\begin{eqnarray}
\label{bb1aa}
F_{S}^{1^-}(\vec{q},\vec{k})= -\frac{m(E_q + E_k)-\vec{q}\cdot\vec{k}}
{2 E_q^2} - \frac{(\vec{q}\cdot\vec{k})^2}{2 E_q^2 \vec{q}^2}
\frac{(E_q + m)}{(E_k + m)},
\end{eqnarray}
coresponding to $0^-$ and $1^-$ mesons, respectively. It is clear that
these coefficients will limit to $m/E$ or it higher order while
$\vec{q}\rightarrow\vec{k}$. On the other hand, however, the static linear
confining potential in momentum space, which behaves as
$G_s(\vec{q}-\vec{k})\propto(\vec{q}-\vec{k})^{-4}$, is strongly
weighted as $\vec{q}\rightarrow \vec{k}$ in Eq.(\ref{e22}) and
(\ref{e22aa}). Because the coefficients will diminish in the relativistic
limit as $\vec{q}\rightarrow \vec{k}$, the strength of the confining
potential would be reduced in turn. This is the reason which leads to the
small spin splittings obtained for heavy-light-quark mesons. However, this
depressing situation would be changed if the retardation is taken into
account. In fact, the covariant form of confinement interaction may take
the form $G_{S} (q , k)\propto [(\vec{q}-\vec{k})^2 - (q_0 - k_0)^2]^{-2}$
and in the on-shell approximation, $q_0^2=m^2 + \vec{q}^2,~k_0^2=m^2 + 
\vec{k}^2$, it becomes
\begin{equation}
G_{S}( q , k ) \propto(-2 m_q^2 + qk - \vec{q}\cdot\vec{k})^{-2}
\end{equation}
in the high energy limit, i.e., $p, k \gg m$. We can see
that with the retardation effect the scalar interaction $G_{S}(q, k)$
is heavily weighted as $\vec{q}$ and $\vec{k}$ are co-linear
($\vec{q}\parallel\vec{k}$), whereas the static linear potential only
peaks at $\vec{q}=\vec{k}$. This indicates that the former is weighted in
a much wider kinematic region than the latter, especially for systems
including light components. As a consequence, the wave functions in
coordinate space would tend to be short ranged and hence the magnitudes of
the wave function at the origin, $\psi(0)$, would increased. Because to
leading order in $1/m^2$ expansion the hyperfine splitting is proportional
to $|\psi(0)|^2$, the Hermitan square of the wave function of meson at the
origin, the splittings would enlarge as well. The above analysis indicates
that in the equal-quark-mass situation the modified effective scalar
interaction will not be weakened too much as $\vec{q}$ and $\vec{k}$
approach to be parallel, and this is just due to the retardation effect. 
In our opinion, the difficulty in the reduced Salpeter equation with the
static scalar confinement is probably due to the
improper treatment that the confining interaction is purely instantaneous
\cite{a1}\cite{ch}. 

In practice, for the constituent quark model, which is essentially used in 
the present work, the equal mass and the on-shell approximation 
maybe not a good simplification for the $q\bar{Q}$ systems, but in any
case the analysis given above is qualitatively correct, which is
supported by the numerical results listed in Table I. And, it is obvious
that the new procedure gives a much better fit to data than the previous
one, in especially the spin splittings. The results obtained
about the mass spectrum, particularly the $s~d$ system, seems still not
fully satisfactory, but it is noted that our results listed are just a
schematic ones. A fine tunning may lead an improvement on them.

In conclusion, we have extented our previous study in clarifying the
problems pointed out by Durand {\it et al.} for the static scalar
confinement in reduced Salpeter equation to the heavy-light-quark systems 
in this paper. The conclusion remains the same that the "intrinsic flaw"
of the Salpeter equation with static scalar confinement could be remedied
to some extent by taking the retardation effect of the confinement into
consideration. A fit for mass spectrum of S-wave heavy-light-quark
systems is obtained by using the scalar linear confinement potential
with retardation and the one-gluon exchange potential. Result shows that a
great improvement is achieved on fitting the data with retardation taken
into account. Although the on-shell approximation may not be a rigorous
treatment here, the qualitative feature of the retardation effect is still
manifest. Nevertheless, it is still premature to assess whether or not the
quark confinement is really represented by the scalar exchange of the form
of $(\vec{p}^2-p_0^2)^{-2}$, as suggested by some authors and being used
here, as the dressed gluon propagator to implement quark confinement.
Therefore, further investigations on this subject are necessary.

\vskip 1cm
\begin{center}
\bf\large\bf{Acknowledgement}
\end{center}
This work was supported in part by the National Natural Science
Foundation of China, the State Education Commission of China, and the
State Commission of Science and Technology of China.

\newpage
\oddsidemargin 0pt
\begin{center}{\large \bf Table I}\end{center}
\begin{center}
\begin{minipage}{18cm}
{Calculated mass spectrum of $ q\bar{Q}$ states using
reduced Salpeter equation with retardation for scalar confinement. The
Experiment data are taken from Ref.\cite{s9}}
\end{minipage}
\end{center}
\oddsidemargin -1.5cm
\vskip 1mm   
\doublerulesep 2.0pt
\begin{tabular}{ccccccc} \hline \hline 
~~State~~  & ~~Quarks~~ & ~~Data (MeV)~~& ~~Fit I$^a$ (MeV)~~& ~~Error
(MeV)~~&~~Fit II$^b$ (MeV)~~& ~~Error (MeV)
\\ \hline
$\bar{B}^0$ & $b\bar{d}$ & 5279 & 5381 & +102& 5258 & -21\\ \hline
$D_s^+$ & $c\bar{s}$ & 1969 & 2097 & +128 & 1946 & -23 \\ \hline
$D_s^{\ast +}$ & $c\bar{s}$ & 2112 & 2148 & +35 & 2094 & -19\\ \hline
$D^+$ & $c\bar{d}$ & 1869 & 1983 & +114 & 1862 & -7 \\ \hline 
$D^{\ast +}$ & $c\bar{d}$ & 2010 & 2010$^*$ & 0 & 2003 & -7\\ \hline 
$\bar{K}^0$ & $s\bar{d}$ & 498 & 743 & + 245 & 652 & +154\\ \hline 
$\bar{K}^{\ast 0}$& $s\bar{d}$ & 892 & 870 & -22 & 898 & +6\\ \hline 
$D_s^{\ast +}-D_s^+$&& 144 & 51 & -93 & 148 & +4\\ \hline
$D^{\ast +}-D^+$&& 141 & 27 & -114 & 141 & 0\\ \hline
$\bar{K}^{\ast 0}-\bar{K}^0$& & 394& 127 & -267 & 244& -150\\ \hline
\hline
\end{tabular}
\vskip 1mm
\noindent
$^a$Results without retardation. $^b$Results with retardation. $^*$Used to
fix the parameters.
\end{document}